\documentstyle[editedvolume,numreferences]{crckapb2}

\begin{opening}
\title{COLLECTIVE BEHAVIOUR AND DIVERSITY IN ECONOMIC COMMUNITIES:
       SOME INSIGHTS FROM AN EVOLUTIONARY GAME}

\author{Vivek S. Borkar}
\institute{Department of Computer Science and Automation \\
Indian Institute of Science \\
Bangalore 560 012, India}
\author{Sanjay Jain}
\institute{Centre for Theoretical Studies \\
Indian Institute of Science \\
Bangalore 560 012, India}
\author{Govindan Rangarajan}
\institute{Department of Mathematics and Centre for Theoretical Studies \\
Indian Institute of Science \\
Bangalore 560 012, India}

\end{opening}

\runningtitle{COLLECTIVE BEHAVIOUR AND DIVERSITY}

\begin{document}

\begin{abstract}
Many complex adaptive systems contain a large diversity of specialized
components. The specialization at the level of the
microscopic degrees of freedom, and diversity at the level of the
system as a whole are phenomena that appear during the course of
evolution of the system. We present a mathematical model to
describe these evolutionary phenomena in economic communities.
The model is a generalization of the replicator equation. The
economic motivation for the model and its relationship with some other
game theoretic models applied to ecology and sociobiology is
discussed. Some results about the attractors of this
dynamical system are described. We argue that while the microscopic
variables -- the agents comprising the community -- act
locally and independently, time evolution produces a collective behaviour
in the system characterized by individual specialization of the
agents as well as global diversity in the community. This 
occurs for generic values of the parameters and initial conditions
provided the community is sufficiently large, and can be viewed
as a kind of self-organization in the system. The context dependence
of acceptable innovations in the community appears naturally in
this framework.
\end{abstract}
\vspace{1cm}  \noindent Preprint No. IISc-CTS-4/98 \\
To appear in the proceedings of the Workshop on Econophysics held
at Budapest, Hungary, July 21-27, 1997.

\newpage
\section{Introduction}

Several complex adaptive systems in the course of their evolution
exhibit the phenomenon that
the individual components comprising the system evolve to perform highly
specialized tasks whereas the system as a whole evolves towards greater 
diversity in terms of the kinds of components it contains or the tasks
that are performed in it.  Here are some examples:

\begin{enumerate}

\item{} Living systems are made of cells, which in turn are made of 
molecules.  Among the various types of molecules are the proteins.  Each
type of protein molecule has evolved to perform a very specific 
task, e.g.,  catalyse a specific reaction in the cell.  At the same time,
during the course of evolution, diverse kinds of protein molecules have 
appeared --
the range of specialized tasks being performed by protein molecules has
increased.  

\item{} In an ecology, species with highly specialized traits appear
(e.g., butterflies with a specific pattern of spots on their wings).
Simultaneously, the ecology evolves to support a diverse variety 
of specialized species.

\item{} Many early human societies (such as hunter-gatherer societies) were
perhaps characterized by the fact that there were relatively few chores
(e.g., hunting, gathering, defending, raising shelter) 
to be performed, and everyone in the community 
performed almost all the chores.  These societies evolved to have 
specialist hunters, tool makers, farmers, carpenters, etc.  Individuals 
specialized, and simultaneously a diverse set of specialists appeared.

\item{} In an economic web, we find firms exploring and occupying increasingly
specialized niches, while the web as a whole supports an increasingly
diverse set of specialists.

\end{enumerate}

In the examples above the systems and their
underlying dynamics are quite different.  But they all share
the twin evolutionary phenomena of individual specialization
and global diversification.  In all these systems, the nonlinear 
interaction among
the components seems to play a crucial role in the manifestation of this
type of behaviour.  For example, in an ecology, the development of highly
specialized traits in a species is a result of its interaction with 
and feedback from other 
species.  In
an economic community, each agent's choices depend upon feedback from 
exchanges  (of goods,
money, etc.)  with other
agents.  Moreover, there is no purposeful global organizing agency
which directs the behaviour of individual components and ordains
them to be specialists. The phenomenon 
happens `spontaneously', arising from the local moves made at
the level of individual components. Similarly diversity also
arises as individuals
capitalize on innovations -- mutations, 
technological innovations, etc. -- which suit them in the existing
context.

In this article, we describe a mathematical model which seems to exhibit 
the above twin evolutionary phenomena.  The (idealized) behaviour of agents 
in economic communities provides the basic motivation of the model.
The model consists of a set of coupled nonlinear ordinary differential 
equations describing the time evolution of the activities of individual
agents. In the next section we motivate and present the model and place it
in the perspective of existing work. 
In section 3
we define more precisely the notions of specialization and 
diversity in the context of the model and outline what type of
behaviour we are looking for. 
Essentially we are seeking attractors of the dynamical model that
have the property of individual specialization and global diversity.
Section 4 states
certain theorems and numerical results for the attractors of the
system and discusses their consequences. The results imply that
under certain conditions that do not destroy genericity 
in parameter space, the
desired attractors (in which the system exhibits individual specialization
and global diversity) exist and have basins of attraction that 
cover the entire configuration space. Thus the evolutionary
phenomena mentioned above occur generically in the model.
In this section we also discuss self-organization and the
emergence of innovations in the model.
Finally, section 5 contains a brief summary.

\section{The model}

The system is a community of $N$ agents labeled by the index 
$\alpha = 1,2, \ldots , N$.  Each agent can perform $s$ strategies or
activities labelled by $i \in S=\{ 1,2, \ldots , s \}$.  At time $t$, agent
$\alpha$ performs strategy $i$ with a probability $p_i^{\alpha}(t)$, 
$\sum_{i=1}^s p_i^{\alpha}(t) = 1$.
The vector ${\bf p}^{\alpha}(t) = (p_1^{\alpha}(t), p_2^{\alpha}(t), \ldots 
, p_s^{\alpha}(t))$ is the mixed strategy profile of agent $\alpha$
at time $t$. 
In particular, if $p_i^{\alpha}(t) = \delta_{ij}$, then
the agent $\alpha$ is said to pursue the pure strategy $j$ or to have
specialized in strategy $j$.  

The vectors ${\bf p}^{\alpha}(t)$ constitute the basic dynamical
variables of the model.
The equation governing their evolution is taken to 
be
\begin{equation}
{\dot p}_i^{\alpha}(t) = p_i^{\alpha}(t) [\sum_{\beta \neq \alpha} 
\sum_{j=1}^s
a_{ij}p_j^{\beta}(t) - \sum_{\beta \neq \alpha} \sum_{i,j=1}^s p_i^{\alpha}(t)
a_{ij}p_j^{\beta}(t)].
\label{2.1}
\end{equation}
Here $a_{ij}$ denotes the $ij$th element of the $s$-dimensional payoff 
matrix $A$.  

This dynamics is motivated as follows.  Each agent is interacting
pairwise with all other agents and receiving a payoff at every 
interaction that depends on the strategy pursued by each of 
the agents during that interaction. Each agent updates her strategy profile
based on the payoffs received, so as to increase her payoff at
subsequent interactions.  In a time $\Delta t$, agent $\alpha$ has 
a total of $m \Delta t$ interactions with every agent ($m$ assumed constant).
If in a particular interaction with agent $\beta$, agent $\alpha$
plays pure strategy $k$ and $\beta$ plays pure strategy $j$, then the
payoff to $\alpha$ is $a_{kj}$ (by the definition
of payoff matrix elements).  Since $\alpha$ plays the strategy
$k$ with probability $p_k^{\alpha}$ and $\beta$ plays the strategy
$j$ with probability $p_j^{\beta}$, the average payoff to $\alpha$ 
from the $m \Delta t$ interactions with $\beta$ is 
$$
m \Delta t \sum_{k,j} p_k^{\alpha}(t) a_{kj} p_j^{\beta}(t).
$$
The average payoff to $\alpha$ from the whole community is
$$
m \Delta t \sum_{\beta \neq \alpha} \sum_{k,j} p_k^{\alpha}(t) a_{kj} 
p_j^{\beta}(t).
$$
This is the second term in the [ ] in Eq. (\ref{2.1}).
We have assumed that $\Delta t$ is large enough for there to be a 
statistically sufficient number of interactions $m \Delta t$ so that
averages make sense.  Yet it is small enough compared to the time
scale at which agents update their strategies so that $p_k^{\alpha}$
can be considered constant during $\Delta t$, i.e., there is a 
separation of time scales between the individual interactions of 
agents (which happen on a short time scale) and the time 
scale over which agents
update their strategy profile (a long time scale).

If agent $\alpha$ were to pursue not the mixed strategy profile
${\bf p}^{\alpha}$ during this interval but instead the pure
strategy $i$, then the payoff received during this period
would have been 
$$
m \Delta t \sum_{\beta \neq \alpha} \sum_{j} a_{ij} p_j^{\beta}(t).
$$
This is the first term in the [ ] in Eq. (\ref{2.1}).  
This quantity depends on
$i$ and for some $i$ will be greater than the average payoff and
for some it will be less than the average payoff.  At the end of
period $\Delta t$, the agent $\alpha$ updates her strategy profile
$p_i^{\alpha}$ to $p_i^{\alpha}+\Delta p_i^{\alpha}$, adding a 
positive weight $\Delta p_i^{\alpha}$ to those 
strategies $i$ that do better than the average and a negative weight 
to those doing worse than the average. $\Delta p_i^{\alpha}/p_i^{\alpha}$
is chosen to be proportional
to the amount by which the pure strategy payoff differs from the
average payoff:
\begin{equation}
\frac{\Delta p_i^{\alpha}}{ p_i^{\alpha}} = c m \Delta t 
 [\sum_{\beta \neq \alpha} \sum_j
a_{ij}p_j^{\beta}(t) - \sum_{\beta \neq \alpha} \sum_{i,j} p_i^{\alpha}(t)
a_{ij}p_j^{\beta}(t)].
\label{2.2}
\end{equation}
Taking the limit $\Delta t \to 0$ and rescaling $t$ by the factor 
$ c m$, we recover Eq. (\ref{2.1}). Therefore the equation 
embodies the statement that at all times, all agents update their
individual mixed strategy profiles so as to increase 
their own payoffs in
the current environment of the strategy profiles of other agents.

The reason why $\Delta p_i^{\alpha}/ p_i^{\alpha}$ and
not just $\Delta p_i^{\alpha}$ appears in the l.h.s. of (\ref{2.2})
is that the dynamics must respect the probability interpretation of 
$p_i^{\alpha}$. If two pure strategies $i$ and $i'$ provide the same payoff to
agent $\alpha$, she must increment them in proportion to their current
strength in her profile. This is needed to ensure that 
${\bf p}^{\alpha}(t)$ remains normalized
at all times, $\sum_{i=1}^s p^{\alpha}(t) = 1$. If we start with 
normalized ${\bf p}^{\alpha}$, the proportionality factor
$p_i^{\alpha}$ on the r.h.s of (\ref{2.1}) ensures that it
remains normalized, since $\sum_{i=1}^s{\dot p}_i^{\alpha}(t)=0$.

Thus, we have a community of $N$ interacting agents, each 
responding to the rest of the environment by updating their own profile
according to the above dynamical equation.  This is a ``non-cooperative 
game''. Agents act on their own (not in concert, {\it per se}) and 
are selfish -- their actions are designed to
increase their own payoff, without consideration for others or the
community as a whole.  Agents also exhibit 
``bounded rationality'' -- they
do not anticipate other agents' future strategies, but merely respond to the
aggregate of the other agents' current strategies.  There is no global
organizing agency at work, the community evolves just through these 
individual actions of the agents.  

Nevertheless, we will argue that the community does exhibit a kind of 
global organization under certain circumstances.  If the community starts
with some 
arbitrary initial condition in which each ${\bf p}^{\alpha}$ at $t=0$ is
specified (each agent starts with some mixed strategy profile which could be
different for different agents)
and evolves according to Eq. (\ref{2.1}), it will settle down
to some attractor of the dynamics.  The organization
referred to above is in the nature of the attractors.  
When the payoff matrix elements satisfy certain 
inequalities, and when the size of the community is larger than a 
certain finite bound that depends on the payoff matrix
(i.e., $N$ is sufficiently large), then we find 
that these attractors are characterized by each individual agent
having specialized to some pure strategy or the other, and at the same time
the community as a whole retaining its full diversity of strategies, i.e.,
every pure strategy is pursued by some agent or the other in the attractor
configuration. Such attractors seem, generically, to be the only
stable attractors of the system under the above conditions. 
Most of the time, we will consider the system with a fixed set of pure
strategies.  At the end, we will mention applications of our results for 
the innovation of new strategies.  
The instability of
attractors in which the community does not have the full diversity of
available strategies provides a mechanism by which new pure strategies,
or innovations, can invade the system.

Before proceeding further, we would like to place this model in the 
perspective of existing work in the subject.  Consider the ``homogeneous
sector", where all agents have the same (but in general mixed) 
strategy profile: $p^{\alpha} = x \ 
\forall \ \alpha$.  Then Eq. (\ref{2.1}) reduces to
\begin{equation}
{\dot x}_i(t) = x_i(t) [ \sum_j a_{ij}x_j(t) - \sum_{k,j} x_k(t) a_{kj}
x_j(t)] (N-1).
\label{2.3}
\end{equation}
The overall factor of $N-1$ can be absorbed in a rescaling of time.  This 
equation is the well known replicator equation \cite{RE}.  It has 
applications in diverse fields such as economics and sociobiology
(where it models evolutionary games)
macromolecular evolution (describing evolution of autocatalytic networks,
in particular the hypercyclic feedback), mathematical ecology (Eq. 
(\ref{2.3}) 
maps onto the Lotka-Volterra equation) and population genetics (where it
is the continuous counterpart of the discrete selection equation).
This system exhibits a great diversity of solutions including fixed points,
limit cycles, heteroclinic cycles, etc.  For more details, see \cite{HS}.
In these applications $i$ labels strategies or species of 
molecules or organisms, depending upon the application. 
$x_i$ represents the fraction of individuals of
type $i$ in a large population, and Eq. (\ref{2.3}) models the 
change of the composition of the {\it population} with time.

Eq. (\ref{2.1}) has been considered as a multi-population generalization
of the replicator
dynamics and has been studied as such in the literature \cite{HS}
\cite{Weibull}.  
The index $\alpha$ now labels populations, e.g., $\alpha=1$ might correspond
to a population of frogs, and $\alpha=2$ to a population of insects.
The idea here is to model the co-evolution of the populations of 
frogs and insects in interaction with each other.
$p^1_i$ now stands for the fraction of the frog population with genotype
$i$ and $p^2_j$ for the fraction of the insect population with genotype
$j$. The indices $i$ and $j$ run over values $s_1$ and $s_2$ respectively
which need not be equal, and now there are two payoff matrices, one
for the frogs, 
$A^1=(a^1_{ij})$, whose matrix element $a^1_{ij}$ equals the payoff
to a frog of type $i$ in an encounter with an insect of type $j$,
and another for the insects, $A^2=(a^2_{ji})$ whose matrix element
$a^2_{ji}$ equals the payoff to the insect in the same encounter.
The dynamics for the two populations is now given by
\begin{eqnarray}
{\dot p}_i^{1}(t) & = & p_i^{1}(t) [\sum_{j=1}^{s_2}
a^1_{ij}p_j^{2}(t) - \ \sum_{k=1}^{s_1} \sum_{j=1}^{s_2} p_k^{1}(t)
a_{kj}p_j^{2}(t)], \nonumber \\
{\dot p}_j^{2}(t) & = & p_j^{2}(t) [\sum_{i=1}^{s_1}
a^2_{ji}p_i^{1}(t) - \ \sum_{k=1}^{s_2} \sum_{i=1}^{s_1} p_k^{2}(t)
a_{ki}p_i^{1}(t)].
\label{2.4}
\end{eqnarray}
This is a so called `bimatrix game' and reduces to Eq. (\ref{2.1}) 
with $N=2$, when $s_1=s_2=s$ and $A_1=A_2=A$.

Our interpretation of  Eq. (\ref{2.1}), presented earlier,
is different from the ``multi-population" interpretation. The
index $\alpha$ labels individuals and not populations.
In the multipopulation interpretation, individuals in each
population are hard-wired to be of some specific genotype, while
the composition of the population is plastic and subject to
selection. For us, the composition of the mixed strategy
profile of each individual is subject to selection.
In the multipopulation interpretation
there is no reason for $A^1$ and $A^2$ to be equal;
frogs and insects {\it are} quite different.
However a single
payoff matrix $A$ is natural in the present context if the
community consists of $N$ {\it identical} agents (identical in that
the payoffs to agents in any interaction depends on the
strategies played in that interaction and not on the
identity of the agents). This allows us to study large
communities (large $N$) without the simultaneous proliferation of
parameters. To our knowledge, the interpretation of
Eq. (\ref{2.1}) as modelling not $N$
populations but a single community of $N$ identical individuals is new.
While we make use of existing mathematical results for Eq. (\ref{2.1}),
the new interpretation prompts us to investigate certain other mathematical 
properties of the model which have not received attention. Since Eq.
(\ref{2.1}) is a
generalization of the replicator dynamics, we will refer to it as the 
generalized replicator dynamics (GRD) whereas Eq. (\ref{2.3})
will be referred to as pure replicator dynamics (PRD).

Note that in (\ref{2.4}) frogs receive payoffs only from insects,
not from other frogs, and insects only from frogs, not from other
insects. This is because the competition among the different genotypes
of frogs happens not directly, but indirectly via their interactions
and competition with insects: the more successful genotypes among frogs
might be the ones (depending upon the payoff matrix)
which do better at capturing insects. Similarly insects do not
compete with each other directly but only with frogs; the insect
population profile evolves because some insect genotypes do better
than others at, say, evading frogs.
A similar justification might be provided for agents in the present
context. A single isolated agent has no competition and hence no
motivation to change her strategy profile.
There is no direct competition among the weights of different pure
strategies within the strategy profile of a single agent; this
competition and consequent evolution arises indirectly because of the
external pressure on the agent from the other agents. A firm that produces
a number of goods
in the economy need not change its production profile if there
are no other producers. But if other producers enter the fray, the firm
may need to change (say, specialize in the production of a few items),
in order to compete effectively. This feature is captured in the model
by the exclusion of the
$\beta=\alpha$ term on the r.h.s. of (\ref{2.1}) -- agents don't
compete with themselves but with other agents. We will see later that
this property is important for the emergence of specialization in
the model.

A well known example is the ``Hawk-dove game" \cite{hawkdove}, in which
there are two pure strategies, ``hawk'' ($i=1$) and 
``dove" ($i=2$).
The payoff matrix elements are $a_{11}=(g-c)/2$, $a_{12}=g$,
$a_{21}=0$, and $a_{22}=g/2$, with (typically) $c > g > 0$. In this 
game individuals interact pairwise and every interaction
is a competition for some resource. In an interaction, a hawk always
escalates and fights, irrespective of what the opponent does. A dove
``displays", but retreats if the opponent escalates. Thus when hawk
meets dove, the dove always retreats and gets zero payoff, while the
hawk gains a payoff $g$ from the resource. When dove meets dove, both
have equal chance of getting the resource or retreating, hence the
average payoff to each party in such an encounter is $g/2$. When hawk
meets hawk, there is a fight, and with equal probability one wins
without injury and gains $g$,
while the other retreats with an injury resulting in a cost $c$. The
average payoff in hawk-hawk encounters to each party is therefore $(g-c)/2$.
It is instructive to contrast the treatment
of this game in PRD and GRD. In the former, there is a large population
of individuals, each hardwired to be pure hawk or pure dove in every
encounter. The fraction of the population that is hawk, $x_1$, and the
fraction that is dove, $x_2=1-x_1$, evolves according to (\ref{2.3})
in response to selection pressure and birth/death processes. The
point $(x_1,x_2)=(g,c-g)/c$ is a stable equilibrium of (\ref{2.3}),
and generically, the population ends up in this attractor, i.e., with
a ratio of hawks to doves being $g/(g-c)$. In GRD, one would have
$N$ agents, each allowed to play both hawk and dove strategy in
an encounter with the respective probabilities $p_1^{\alpha}$ and
$p_2^{\alpha}$. It is not obvious that each agent will end up specializing
in a pure hawk or pure dove strategy, but that is what does happen.
A consequence of one of the theorems to be described later is that
the only stable attractor of this GRD is a configuration where
agents tend to distribute themselves in a pure hawk or pure dove
strategy roughly in the ratio $g/(c-g)$ for finite $N$, and exactly
in this ratio as $N \rightarrow \infty$. Thus individual specialization,
which was true by assumption in PRD, is a dynamical outcome in GRD.
Moreover, while individuals specialize in their self interest to
some pure strategy or the other depending upon their initial conditions,
collectively the community seems to obey some global constraints.

\section{Definitions and Notation}

Consider
$$
J=\{ {\bf x} = (x_1, x_2, \ldots, x_s) \in R^s | \sum_{i=1}^s x_i = 1,
x_i \geq 0 \},
$$
which is the simplex of $s$-dimensional probability vectors.  $J$ is the
full configuration space of PRD dynamics and is invariant under it.
The configuration space for GRD is $J^N$, the $N$-fold 
product.  A generic point of $J^N$ is $p=({\bf p}^1, {\bf p}^2, 
\ldots , {\bf p}^N)$, each
${\bf p}^{\alpha}$ being an $s$-dimensional probability vector 
${\bf p}^{\alpha}(t) = (p_1^{\alpha}(t), p_2^{\alpha}(t), \ldots 
,p_s^{\alpha}(t))$ belonging to $J^{(\alpha)}$ (the latter being
a copy of the simplex $J$ corresponding to
agent $\alpha$).  

A point ${\bf x} \in J$ such that $x_i = \delta_{ij}$ for some $j$ is called the
$j$th corner of $J$.  
If an agent $\alpha$ has specialized to the pure strategy $j$, then 
$p_i^{\alpha} = \delta_{ij}$, i.e., ${\bf p}^{\alpha}$ has gone to the $j$th
corner of $J^{(\alpha)}$.  If every agent has specialized to some 
strategy or the other, the corresponding point in $J^N$ will be called a
{\it corner} of $J^N$ and we say that the community is {\it fully
specialized}.  Note
that every corner of $J^N$ is an equilibrium point of GRD, since the
r.h.s. of (\ref{2.1}) vanishes.
Hence we refer to
corners as corner equilibrium points (CEPs).

A CEP can be characterized by an $s$-vector of non-negative integers 
${\bf n}=(n_1, n_2, \ldots , n_s)$ where $n_i$ denotes the number of agents
pursuing the pure strategy $i$ at CEP.  There can be many CEPs with the same
${\bf n}$ vector.  These would differ only in the identity of the agents at
various corners. In this article we will ignore the differences between
such corners and characterize a CEP by its
${\bf n}$-vector alone, since the agents are identical and differ
only in their strategy profile.

Consider the following subset of $J^N$:
$F_k \equiv \{ p \in J^N | p_k^{\alpha} = 0 \ \forall \ \alpha \}$ for some
fixed $k \in S$. By definition, at a point in $F_k$, every agent has
opted out of strategy $k$. $F_k$ is also invariant under (\ref{2.1}),
i.e., if $p_k^{\alpha}$ is zero at some time, it remains zero. At the
``face" $F_k$, strategy $k$ therefore becomes extinct from the population,
and we say that the full diversity of strategies is lost. As long as
the system is not in some $F_k$, we  say that the community exhibits
the {\it ``full diversity"} of strategies. 
Note that the word `diversity', as used here, does not
stand for variability among agents, but to indicate that all strategies
are supported. For example we can have no variation but full diversity
at points $p$ where ${\bf p}^{\alpha} = {\bf c} \ \forall \ \alpha$
and none of the components of ${\bf c}$ are zero. This is a ``homogeneous"
point, since all agents are doing the same thing.

If a CEP is such that $n_i \neq 0$ for all $i$, i.e.,
each strategy is played by at least one agent at the CEP, we will refer to it
as a {\it fully diversified} CEP or FDCEP. If one or more $n_i$ is zero,
the full diversity of strategies is lost and such CEPs are called 
non-FDCEPs.

We are interested in studying the circumstances under which FDCEPs are 
the preferred attractors of the dynamics, for then, individual specialization
and global diversity will arise dynamically in the community.  If it 
happens that the FDCEPs are attractors and their basins of attraction cover
most of $J^N$ (all of $J^N$ except a set of lower dimension),
then for generic initial conditions
the community
is bound to end up in an FDCEP, which means that it will exhibit
individual specialization and as well as global diversity. 

\section{Results}

In this section we discuss some results concerning attractors of GRD.
The proofs of the theorems are omitted here; these and further
results can be found in
\cite{BJR1,BJR2}. We will discuss the significance of these
results for specialization and diversity in GRD.

\subsection{Interior Equilibrium Points}

An equilibrium point of GRD is called an interior equilibrium point (IEP) if 
none of the $p_i^{\alpha}$ is zero. We have the following theorem:

{\bf Theorem 1:} There is at most one isolated IEP.  If there is one, it
is homogeneous and is given by $p_i^{\alpha} = x_i \ \ \forall \ \ \alpha, i$
where $x_i \equiv u_i/{\rm det} B$, $u_i$ is the cofactor of $B_{0i}$, and
$B$ is the $(s+1) \times (s+1)$ matrix (whose rows and columns are
labelled by the indices $0,1,2, \ldots, s$)
\begin{equation}
B \equiv
\left( \begin{array}{ccccc}
0&1&1&\cdots&1 \\
-1&&&&         \\
-1&&A&&        \\
\vdots &&&&    \\
-1&&&&
\end{array} \right). \label{4.1}
\end{equation}
A necessary and sufficient condition for an isolated IEP to exist is given by
\begin{description}
\item{} {\bf A1}: \ \ \ $u_i \neq 0 \ \ \forall \ \ i$, and all $u_i$ have the
same sign.
\end{description}

PRD also has an isolated IEP if condition {\bf A1} holds, which
is then unique and given by the same $x_i$ as given above for GRD.  Further,
note that at the IEP in GRD, the system exhibits full diversity since no
strategy is opted out of by any agent.  However, there is no specialization.
The above formula for the IEP in particular yields the point
$(x_1,x_2)=(g, c-g)/c$
for the hawk-dove game.

\subsection{Specialization}

For generic payoff matrices $A$, generic initial conditions, and sufficiently
large $N$, we find that the system flows into a corner of $J^N$.  Thus
specialization is a generic outcome of the dynamics.  This observation is 
based on the following facts:  

\begin{enumerate}

\item{} {\bf Theorem 2:} \cite{HS,Weibull} Any compact set in the interior of $J^N$ or
the relative interior of any face cannot be asymptotically stable.  
An equilibrium point is asymptotically stable if and only if it is a 
strict Nash equilibrium.
(A strict Nash equilibrium
is a point $p$ such that at this point if any single agent unilaterally
changes her strategy -- unilaterally means that all other agents
remain where they are -- then her payoff strictly decreases.)

\item{} {\bf Theorem 3:}  Every asymptotically stable attractor must
contain at least one corner equilibrium.

\item{} {\bf Numerical Work:} The GRD equation for $s=3$ was numerically 
integrated using Runge-Kutta method of fourth order.  
We randomly generated ten
$3 \times 3$ payoff matrices and numerically integrated
the GRD equations for long times for each payoff matrix with ten
randomly chosen initial conditions. When this was done with $N=5$,
in 90 out of the 100 cases the dynamics converged to a corner. The
remaining 10 cases (all corresponding to a single payoff matrix) 
converged to a heteroclinic cycle. (In these 10 cases
the system cycled between regions
close to a few corners, moving rapidly between these regions, and
at every successive cycle
spending increasing amounts of time near the
corners and coming closer to them.) When $N$ was increased to 10
for the same ten payoff matrices studied above, all 100 cases 
converged to a corner. This suggests that the typically, the stable
attractors are corners or heteroclinic cycles, with corners becoming
overwhelmingly more likely at larger $N$.

\end{enumerate}

Why are corners the preferred attractors of this dynamics?
We give here an intuitive argument, which, though not rigorous
or complete,
provides some insight. (For rigorous arguments, refer to the
proofs of the above mentioned theorems.)
Recall
that according to the dynamics each agent updates her strategy
profile to increase her payoff in the current environment. Pick
an agent $\alpha$. Her payoff at any point is
$P^{\alpha} = \sum_k p_k^{\alpha} c_k^{\alpha}$
where $c_k^{\alpha} \equiv \sum_{\beta \neq \alpha} a_{kj} p_j^{\beta}$.
Given a set of $s$ numbers $c_k^{\alpha}$ for a fixed $\alpha$,
generically one of them will be the largest. Let the largest one
be $c_l^{\alpha}$ (for some particular $l$). Then it is clear
that since the payoff $P^{\alpha}$ is linear in $p_k^{\alpha}$, the
choice $p_k^{\alpha}=\delta_{kl}$ will maximize it.
Thus, as long as the index of the largest of the
$c_k^{\alpha}$ remains $k=l$, the agent $\alpha$ will
move towards the pure strategy $l$. This argument can be made for
any agent. Thus every agent is, at any time, moving towards some
pure strategy. In this argument it is crucial that $c_k^{\alpha}$
is independent of ${\bf p}^{\alpha}$ (which it is because of the
exclusion of the $\beta=\alpha$ term in the payoff to $\alpha$).
If it were not, then the nonlinear dependence of $P^{\alpha}$
on ${\bf p}^{\alpha}$ would have invalidated the argument
(as is the case in PRD, where the analogous quantity 
$\sum_{k,j}x_k a_{kj} x_j$
is quadratic in the $x_i$, and corners are not the generic
attractors).

This argument also sheds some light on why
heteroclinic cycles could be attractors. The point is that
$c_k^{\alpha}$ are not constants, but depend upon the strategy profiles
of agents other than $\alpha$. If the change in these profiles
causes some other $c_k^{\alpha}$ (for some $k=l'$, different from $l$)
to overtake $c_l^{\alpha}$, then from that time onwards, agent $\alpha$
will have to change track and move towards pure strategy $l'$ rather than
$l$.

In particular the above theorems mean that the IEP is always unstable
in GRD.

\subsection{Collective behaviour}

We have seen above that the GRD flows to corners generically.  The next
step is to determine {\it which} corners the dynamics flows to.  
For the moment we restrict ourselves to FDCEPs.  Characterizing an FDCEP
by its ${\bf n}$ vector (described in the previous section), we have the
following result:

{\bf Theorem 4:} Let ${\bf n}$ and ${\bf n}'$ be any two asymptotically stable
FDCEPs with $N \ge s$.  If condition {\bf A1} holds then all components
of ${\bf n-n}'$ are bounded by a function of the payoff matrix $A$ alone (and
not of $N$).  Further,
$$
\lim_{N \to \infty} {n_i\over N} = x_i.
$$

Thus, out of a large number (of order $N^{s-1}$) of FDCEPs all of which
are equilibria for GRD, only a few are stable.  Further, even though
the agents are acting individually and selfishly and going to corners
(specializing), the system as a whole retains a memory of the unique interior
equilibrium point (which is guaranteed to exist under the conditions of
the above theorem) and tunes the ratios $n_i/N$ such that they are close
to the
IEP $x_i$ values.

One can give a physical or ``economic" interpretation of this
collective behaviour. A stable equilibrium
is by theorem 2 a strict Nash equilibrium. Thus it cannot be
advantageous for any agent to switch her pure strategy unilaterally.
This means that all agents must receive more or less the same
payoff. More precisely, since a switch of strategy by a single
agent causes changes of $O(1)$ in the payoffs to other agents,
at a strict Nash equilibrium it must be the case that differences
of payoffs among agents could not be larger than $O(1)$,
since otherwise it would be possible for some agent to
make an advantageous switch without affecting others. Thus
stability is achieved only at those equilibria at which differences
in payoff among agents are a very small fraction of 
the total payoff to any agent, which is $O(N)$. This requirement
of ``near-equality" of payoffs narrows down the set of
stable equilibria considerably. As to why the ratio $n_i/N$ gets
tuned to be close to the IEP, we remark that in both PRD and GRD,
the IEP is characterized by exactly equal payoffs to all
strategies.

The above remarks also help explain some of our numerical results.  When we 
increased the number of agents from 5 to 10, we found that all 
cases converged to corners.  This is because as $N$ increases, the
ratios $n_i/N$ can reproduce the $x_i$ values corresponding to 
IEP more accurately and thereby achieve the near-equality
of payoffs required for the existence of a strict Nash equilibrium.

\subsection{Diversity and self-organization}

We have seen above that among the FDCEPs, only a very small subset
can be asymptotically
stable. Now we consider non-FDCEPs. It could happen that along with
an FDCEP, some non-FDCEPs are also stable. In that case, if the community
starts in the basin of attraction of a non-FDCEP, it would eventually
lose its diversity. We would like to eliminate such attractors of
the dynamics. It turns out that this can be achieved by imposing
certain inequalities on the payoff matrix elements.
Consider the following condition:

{\bf A2:} The payoff matrix is diagonally subdominant, i.e.,
$a_{ii} < a_{ji} \forall j \ne i$.

{\bf Theorem 5:} For $s=2$, if {\bf A2} holds, then all non-FDCEPs are
unstable for $N \ge 2$.  For $s=3$, if {\bf A1}, {\bf A2} hold,
then there exists a positive number $N_0$
depending on $A$, such that all non-FDCEPs are unstable for $N > N_0$.

The condition {\bf A2} means that each pure strategy gives more
payoff to other strategies than itself, clearly a tendency that
would support diversity. It is interesting that the model also has the
desirable feature that larger communities favour diversity.

It will be useful to have conditions for higher $s$ also which make
all the non-FDCEPs unstable. Partial results in this direction are
contained in \cite{BJR2}. We expect that for higher $s$ the condition
of sufficiently large $N$ and further inequalities on the payoff matrix
elements would ensure the instability of non-FDCEP.

We now discuss the behaviour of the system when such conditions hold.
Notice that since these conditions are inequalities on the payoff matrix
elements (and not equalities), the behaviour of the system is
structurally stable or generic, i.e., is not destroyed by a small
perturbation of the parameters. From the evidence presented in section
5.2, the system is expected to go to a corner with generic initial
conditions. By theorem 5 (and its
generalizations to higher $s$), this corner cannot be a non-FDCEP.
Hence it must be an FDCEP. But then theorem 4 applies and tells us
that it must be a very specific corner. At this corner the number of
agents $n_i$ pursuing the pure strategy $i$ is fine tuned to a value
close to $Nx_i$ with $x_i$ determined by the payoff matrix via
Theorem 1, and all agents receive the same $O(N)$ payoff upto
differences of $O(1)$. The final state is fine-tuned
but robust in that it arises 
without fine-tuning
the parameters or the initial state. In this sense
(of spontaneous dynamical fine-tuning) the system
exhibits self-organization, albeit without any obvious critical
behaviour. Furthermore, this self organization is of the kind that
we were originally seeking, namely, in which the community exhibits
individual specialization
and global diversity.

\subsection{Innovations}

So far we have considered strategy spaces of a fixed size $s$. However,
the growth of diversity in the systems mentioned in the introduction
has to do with the appearance of new strategies and disappearance of
some old strategies. We now discuss how the above considerations of
the instability of non-FDCEPs are relevant for the generation of
innovations. As an example consider a community of $N$ agents which has initially
only two strategies and a payoff matrix 
$A$ satisfying condition {\bf A2}.  Then all non-FDCEPs are unstable and
only those FDCEPs with ${\bf n}=(n_1,n_2)$ such that the ratios $n_i/N$ close to 
$x_i$ of the IEP are stable. Let us assume that the system has settled
into such a state.
Now, assume that a new, third strategy arises which enlarges the $2 \times 2$
matrix $A$ into a 
$3 \times 3$ matrix $A'$
that contains $A$ as a submatrix. The third row and column of $A'$
represent the relationship of the new strategy with respect to the old --
how much payoff it gives to them and receives from them. At the time
this strategy arises, the system is in the state $(n_1,n_2,0)$
since the third strategy (being new) is as yet unpopulated.  Note that
this state is a non-FDCEP for $s=3$.  Now if
$A'$ is such that it satisfies the conditions {\bf A1, A2}
in Theorem 5, and $N$ is sufficiently large, then the above
state is unstable.  Therefore, any small perturbation in which the
agents start exploring the new strategy ever so slightly will
destroy the old state and take the system to
a new stable state which must be an FDCEP with 3 strategies.
By Theorem 4 this state will be $(n'_1,n'_2,n'_3)$ with
$$
\frac{1}{N} (n'_1,n'_2,n'_3) \approx (x'_1,x'_2,x'_3)
$$
where $x'_i$ are the components of the IEP corrsponding to the
payoff matrix $A'$ and are all non-zero since $A'$ satisfies {\bf A1}.
Thus the innovation has destabilized the previous state of the
system and brought it to a new state where a finite fraction of the population
has adopted strategy 3.  In such a case, we say that the innovation 
has been ``accepted'' by the community.
Note that the only requirements for this to happen is that the elements of the
new row and column in the payoff matrix satisfy certain inequalities
with respect to the existing matrix elements (contained in the conditions
{\bf A1, A2} to be satified by $A'$) and that the community be sufficiently
large. This tells us what properties a new strategy should have
{\it in the context of already existing activities} in order 
for it to be ``accepted" by the
community. Thus the model suggests a natural mechanism for the
emergence of context dependent innovations in the community.

\section{Conclusions}
To summarize, 
Generalized Replicator Dynamics, eq. (\ref{2.1}),
is a nonlinear dynamical model of learning for a
community of $N$ mutually interacting agents with the following features:
\begin{enumerate}

\item{} Each agent is selfish and exhibits bounded rationality.

\item{} This is a non-cooperative game and there is no global organizing
agency at work. It is in general a non-hamiltonian system. 

\item{} Specialization of individual agents to pure strategies is a generic
outcome of the dynamics.

\item{} Under certain generic conditions on the payoff matrix parameters
the agents exhibit a collective behaviour,
and for sufficiently large $N$, the community exhibits diversity and
self-organization.

\item{} A `good' innovation (one that satisfies conditions {\bf A1, A2},
etc., with respect to the exisiting strategies) makes the society unstable
and evolve until the innovation is accepted.

\end{enumerate}

It is noteworthy that in this dynamical system, order is generated
at large $N$ (unlike the systems where increasing the number of
degrees of freedom makes the system less orderly, in some sense).
This order is not the usual statistical mechanical kind of order, the order
of appropriately defined macroscopic variables, but
an order in the original dynamical variables themselves.  However,
in another sense, this order is also statistical since we do not
know which pure strategy an individual agent follows.  We only know
about the fraction of agents pursuing a given strategy.

\vspace{2mm}
\noindent{\bf Acknowledgement}:
SJ acknowledges support from the Jawaharlal Nehru Centre for
Advanced Scientific Research, Bangalore, and the Associateship
of the International Centre for Theoretical Physics, Trieste.

\vspace{2mm}
\noindent{\bf Email addresses}: \\
borkar@csa.iisc.ernet.in, 
jain@cts.iisc.ernet.in, rangaraj@math.iisc.ernet.in


{\small

}
\end{document}